\documentclass[intlimits,twoside,a4paper]{article}

\usepackage{graphicx,epstopdf}
\usepackage{booktabs}
\usepackage[cp1251]{inputenc}

\usepackage[eqsecnum]{cmpj3}

\issue{2017}{20}{2}{23601}
\doinumber{10.5488/CMP.20.23601}

\title[Optical, elastic and thermodynamic properties of TiN: A first-principles study]%
{First-principles study of optical, elastic anisotropic and
thermodynamic properties of TiN under high temperature and high pressure%
}
\author[R. Yang, C. Zhu, Q. Wei, K. Xiao, Z. Du]{R. Yang\refaddr{label1}\thanks{E-mail: yrk18687@163.com}\,, C. Zhu\refaddr{label1}, Q. Wei\refaddr{label1}, K. Xiao\refaddr{label1}, Z. Du\refaddr{label2}}
\addresses{
\addr{label1} School of Physics and Optoelectronic Engineering, Xidian University, Xi'an, Shaanxi 710071, PR China
\addr{label2} National Supercomputing Center in Shenzhen, Shenzhen 518055, PR China
}

\date{Received November 24, 2016, in final form April 18, 2017}
\begin{document}

\maketitle

\begin{abstract}
The optical, elastic anisotropic and thermodynamic properties of TiN in the
NaCl (B1) structure are analyzed in detail in the temperature range from 0
to 2000~K and the pressure range from 0 to 20~GPa. From the calculated
dielectric constants, a first order isostructural phase transition between
29 and 30~GPa is found for TiN. The absorption spectra exhibit high values
ranging from the far infrared region to the ultra-violet one. The anisotropy
value of Young's modulus of TiN is smaller than that of c-BN at 0~GPa and
the anisotropy of TiN clearly increases with an increase of pressure. The
effects of pressure and temperature on the bulk modulus, Gr\"{u}neisen
parameter, Gibbs free energy, and Debye temperature are significant. The
Gr\"{u}neisen parameter of TiN is much larger than that of c-BN. At
temperatures below 1000~K, TiN's heat capacity is much larger than that of
c-BN.
\keywords TiN, optical properties, elastic anisotropy,
thermodynamic properties, first-principles
\pacs 61.82.Bg, 62.20.dc, 71.20.Be, 71.15.Mb
\end{abstract}

\section{Introduction}
As one of transition-metal nitrides, titanium nitride (TiN) was first
separated by Story-Maskelyne from a meteorite. It crystallizes into the
well-known rock-salt structure. It is one of the important metal cutting
tools and coating materials for surface protection due to its high melting
temperature, extreme hardness and fine corrosion resistance \cite{1,2,3,4,5,6}. Its
ceramics have a good resistance to corrosion by liquid steel for some steel
making processes \cite{7}. It is also used for diffusion barriers,
superconducting devices, and energy-saving coatings for windows due to their
strong infrared reflection \cite{8}. The high-pressure and high-temperature
studies on various materials are very important both from basic and
applied point of view. Up to now, some elastic and thermodynamic properties
of TiN in the NaCl (B1) structure have been calculated. Some mechanical
properties of TiN were studied in  \cite{3, 4, 9}. Liu et~al. studied the
pressure and temperature dependences of the relative volume, the Debye
temperature, heat capacity and thermal expansion for TiN \cite{10}. An anomalous
volume behavior of TiN between 7.0 and 11.0~GPa was found \cite{11}, which was
interpreted as a first order isostructural phase transition. Recently, a
phase transformation between 35~GPa and 57~GPa was shown by the volume
collapse of TiN \cite{12}.

The effect of the first order isostructural phase transition on the optical
properties of TiN has not been reported so far. Thus, the optical
properties are analyzed in our works by first-principles calculations.
Moreover, the relationship between Young's modulus anisotropy and pressure is
discussed. Several thermodynamic properties of TiN are analyzed in detail at
higher temperature and pressure. To the best of our knowledge, these
properties have not been discussed in previous literature. As one of the
groups III-nitrides semiconductor compounds, cubic boron nitride (c-BN)
exhibits many fascinating mechanical and thermal properties, such as,
extreme hardness, high melting point, high thermal conductivity, and so on.
Thus, some properties of c-BN are also calculated for comparison.

\section{Computational methods}
Structural optimization and prediction of properties are performed using the
density functional theory~(DFT) with the generalized gradient approximation~(GGA) parameterized by Perdew, Burke and Ernzerrof~(PBE) as implemented in
the Cambridge Serial Total Energy Package (CASTEP) code \cite{13}. The optical
properties are analyzed using PBE0 hybrid functional. The cut-off energy
determines the numbers of plane waves, while the number of special k-points is
used for the Brillouin zone (BZ) division \cite{14}. Convergence tests prove that
the Brillouin zone sampling and the kinetic energy cut-off are reliable to
guarantee excellent convergence \cite{15}. For the geometry optimization, the
self-consistent field tolerance threshold is taken as $ 5\times 10^{-7}$~eV/atom. The convergent value of the total energy difference is less than
$ 5 \times 10^{-6} $~eV/atom. The maximum Hellmann-Feynman force is given as
0.01~eV/{\AA}. The maximum stress is taken as less than 0.02~GPa. The maximum
atom displacement is smaller than $5\times 10^{-3} $~{\AA}. The energy
cut-off values are 580~eV for TiN and 550~eV for c-BN. The k-points in the
first Brillouin zone are $14 \times 14 \times 14$ for TiN and c-BN. The
elastic constants are calculated by using the stress-strain methods. In the
strain calculation, the number of steps is set as 6. Maximum strain
amplitude is taken as 0.003. In the calculations of the optical properties,
the norm-conserving pseudo-potentials have been applied to all the
ion-electron interactions. The scissors operator is taken as 0.5~eV. The
instrumental smearing (which~describes the Gaussian broadening to be used
for calculating the dielectric function) is set to 0.2~eV and the incident
light direction is (100).

The optical properties are analyzed for TiN, based on GGA-PBE and PBE0
calculations. Generally, the optical properties of systems are evaluated by
the complex dielectric function, which is dependent on frequency and is as
follows:
\begin{equation}
\label{eq1}
\varepsilon (\omega )=\varepsilon_{1} (\omega )+\ri\varepsilon_{2}
(\omega).
\end{equation}
The complex dielectric function is mainly connected to the electronic
structures. The imaginary part of the dielectric function, $ \varepsilon_2 (\omega ) $, is calculated based on the momentum matrix elements
between the occupied and unoccupied wave functions, it is as follows \cite{16}:
\begin{equation}
\label{eq2}
\varepsilon_{2} (\omega )=\frac{Ve^{2}}{2\piup\hbar m^{2}\omega^{2}}\int
{\rd^{3}k\sum\limits_{n,{n}'} {\left| {\left\langle {kn\left| p \right|k{n}'}
\right\rangle } \right|^{2}f(kn)[1-f(k{n}')]\delta (E_{kn} -E_{k{n}'}
-\hbar \omega )} },
\end{equation}
where $ V $ is the unit cell volume, $ e $ is electronic charge, $ n $ and $ n' $ are the initial
state and final state band indexes, respectively, $ p $ denotes the momentum
operator,  $|kn\rangle $ expresses a crystal wave function, $ f(kn) $ represents the Fermi
distribution function, and the energy of the incident photon is $ \hbar\omega $. The real part, $ \varepsilon_{1} (\omega ) $, is given by
Kramers-Kroning relationship:
\begin{equation}
\label{eq3}
\varepsilon_{1} (\omega )=1+\frac{2}{\piup }M\int\limits_0^\infty
{\frac{\varepsilon_{2} ({\omega }'){\omega }'}{{\omega }'^{2}-\omega^{2}}}
\rd{\omega }',
\end{equation}
where $ M $ is the principal value of the integral. The real part of dielectric
function is in relation to the electric polarization characteristics of the
material. Optical reflectivity, $ R(\omega ) $, and absorption coefficient,
$ I(\omega ) $, can be calculated based on the complex dielectric
function $ \varepsilon (\omega ) $, respectively. These expressions are shown
as follows \cite{17, 18}:
\begin{align}
\label{eq4}
R(\omega )&=\frac{(n-1)^{2}+k^{2}}{(n+1)^{2}+k^{2}}\,,\\
\label{eq5}
I(\omega )&=\frac{2\omega k(\omega )}{c}\,.
\end{align}

An illustrative way of describing the elastic anisotropy is a
three-dimensional surface representation. It shows the variation of the elastic
modulus with crystallographic direction. This directional dependence of the
Young's modulus $ E $ for a cubic crystal is given by \cite{19}:
\begin{equation}
\label{eq6}
\frac{1}{E_{\text{Cubic}} }=s_{11} -2\left(s_{11} -s_{12} -\frac{1}{2}s_{44} \right)\left(l_{1}^{2}
l_{2}^{2} +l_{2}^{2} l_{3}^{2} +l_{3}^{2} l_{1}^{2} \right).
\end{equation}

Based on the optimized crystal structures, the thermodynamic properties for
TiN under high temperature and high pressure are calculated by means of the
quasi-harmonic Debye model, in which the phonon effect is considered and the
results are achieved by GIBBS program \cite{20}. The temperature is included in
the quasi-harmonic Debye approximation. In fact, thermodynamic
properties of the crystal can be obtained by treating the lattice vibrations
as a quantized phonon. The model has been successfully used to predict the
thermodynamic properties of some materials \cite{21,22,23}. In the quasi-harmonic
Debye model, the non-equilibrium Gibbs function $ G^{\ast }(V;P,T) $ takes the
form of \cite{24}
\begin{equation}
\label{eq7}
G^{\ast }(V;P,T)=E(V)+PV+A_{\text{Vib}} \big(\Theta_{\text D} (V);T\big),
\end{equation}
where $ E(V) $ is the total energy per unit cell, $PV$ corresponds to the constant
hydrostatic pressure condition, $ \Theta_{\text D} (V) $ is the Debye temperature,
$ T $ is the absolute temperature and $ A_{\text{Vib}} $ is the vibrational contribution,
that can be written as follows:
\begin{equation}
\label{eq8}
A_{\text{Vib}} (\Theta_{\text D}; T)=nkT\left[\frac{9}{8}\frac{\Theta_{\text D} }{T}+3\ln
\left(1-\re^{-\Theta_{\text D} /T}\right)-D(\Theta_{\text D} /T)\right],
\end{equation}
where $ n $ is the number of atoms in a chemical formula, $ D(\Theta_{\text D} /T) $
represents the Debye integral with Poisson ratio $ \sigma $. The $ \Theta_{\text D}
$ is expressed by
\begin{equation}
\label{eq9}
\Theta_{\text D} =\frac{\hbar }{k}\left(6\piup^{2}V^{1/2}n\right)^{1/3}f(\sigma
)\sqrt{\frac{B_{S} }{M}}\,,
\end{equation}
where $ M $ is the molecular mass per unit cell, $ B_{S} $ is the adiabatic bulk
modulus, which is equal to the isothermal bulk modulus $ B_{T} $ in the Debye
model, leading to the following equation:
\begin{equation}
\label{eq10}
B_{S} =B_{T} =V\frac{\rd^{2}E(V)}{\rd V^{2}}\,,
\end{equation}
where $ E $ is the total energy of the crystal at 0~K.

For the non-equilibrium Gibbs function, $ G^{\ast }(V;P,T) $, can be minimized
with respect to volume $ V $. There is
\begin{equation}
\label{eq11}
\left[\frac{\partial G^{\ast }(V;P,T)}{\partial V}\right]_{P,T} =0.
\end{equation}
By solving the equation \eqref{eq11} with respect to $ V $, a thermal equation of state
can be gotten. The heat capacity, $ C_{V} $, and thermal expansion coefficient,
$ \alpha $, are given as follows:
\begin{align}
\label{eq12}
C_{V} &=3nk\left[4D(\Theta /T)-\frac{3\Theta /T}{\re^{\Theta /T}-1}\right],\\
\label{eq13}
\alpha &=\frac{\gamma C_{V} }{B_{T} V}\,.
\end{align}

Anharmonic effect of the vibrating lattice is usually described in terms of
Gr\"{u}neisen parameter, $ \gamma $, which can be defined as
\begin{equation}
\label{eq14}
\gamma =-\frac{\rd \ln \Theta (V)}{\rd\ln V}\,.
\end{equation}
The heat capacity $ C_{P} $ is expressed as
\begin{equation}
\label{eq15}
C_{P} =C_{V} (1+\alpha \gamma T).
\end{equation}

The entropy, whose change between states is defined as the integral of the
ratio of the reversible heat transfer to the absolute temperature, is a
measure of the state of disorder of the system. Entropy $(S_{\text{Vib}})$ in the
quasi-harmonic Debye model is given by \cite{20}
\begin{equation}
\label{eq16}
S_{\text{Vib}} =nk\left[4D(\Theta /T)-3\ln \left(1-\re^{-\Theta /T}\right)\right].
\end{equation}

\section{Results and discussion}

\subsection{Optical properties}

\begin{figure}[!b]
\centerline{\includegraphics[width=0.5\textwidth]{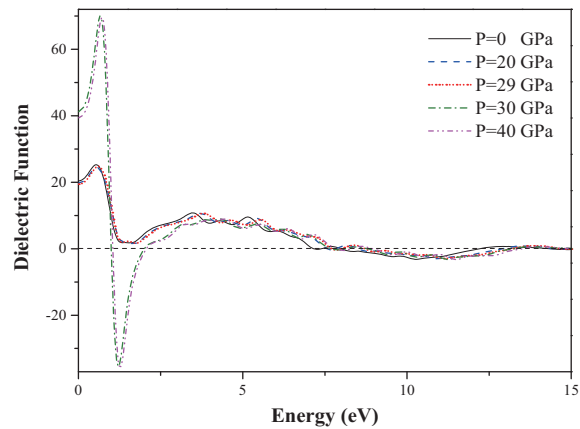}}
\caption{(Color online) Real part of the dielectric function calculated by GGA for TiN.}
\label{fig1}
\end{figure}

Dielectric function, being the bridge between the microscopic physical process
and the electronic structure of solid by electronic inter-band transition,
reflects the energy band structure of solid materials and other optical
information. Figure~\ref{fig1} presents the real part of the calculated dielectric
function from 0 to 40~GPa for TiN. The calculated static dielectric
constants $ \varepsilon (0) $ are 20.4, 19.8, 19.4, 41.2 and 39.4 when the
pressures are 0, 20, 29, 30 and 40~GPa, respectively. The $ \varepsilon (0) $
decreases with an increase of pressure from 0 to 29~GPa, but it shows an
anomalous behavior between 29 and 30~GPa, which is interpreted as a first
order isostructural phase transition. The real part of dielectric function
enhances with increasing photon energy and gets to the highest values at
about 0.6~eV, and afterwards it is a damped oscillation. In the following work,
we only study the properties of TiN from 0 to 20~GPa.
\begin{figure}[!b]
\centerline{\includegraphics[width=\textwidth]{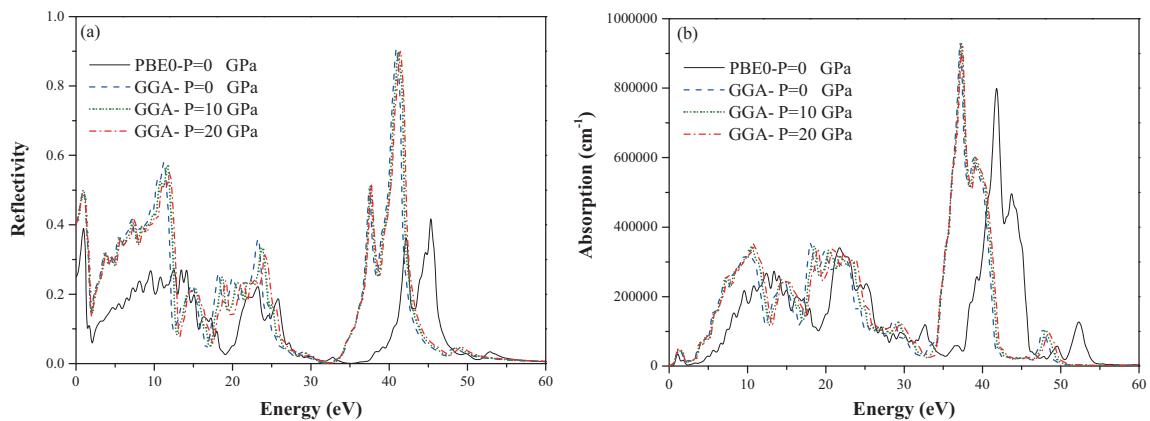}}
\caption{(Color online) (a) Reflectivity $ R(\omega ) $ and (b) absorption coefficient
$ I(\omega ) $ by GGA-PBE and PBE0 from 0 to 20~GPa for TiN.}
\label{fig2}
\end{figure}

The calculated results in figure~\ref{fig2}~(a) and~(b) reveal that TiN possesses a broad
frequency response. It has strong reflectivity and absorption both in the
low-frequency and high-frequency region. The GGA results are plotted and
compared with the PBE0 results. The reflectivity spectra of TiN are
characterized by six main peaks. The GGA results show that the peaks are
located at 1.0, 11.2, 20.0, 23.3, 37.5 and 40.9~eV at zero pressure, while
the PBE0 results show they are located at 1.0, 14.1, 23.3, 25.8, 42.2 and
45.4~eV.

The optical absorption spectra are directly connected to the imaginary part
of the refractive index, which represents the fraction of the energy lost
when a light wave passes through materials. When materials absorb energy,
their electrons jump up to higher energy levels. Therefore, the interband
transitions in different energy regions are responsible for the pronounced
peaks of absorption spectra. The absorption spectra exhibit high values
ranging from the far infrared region to the ultra-violet one. One can notice that
the absorption peaks are positioned at 10.20, 14.6, 21.3 and 38.9~eV
according to GGA, and 15.0, 21.8 and 41.8~eV according to PBE0, in figure~\ref{fig2}~(b).

\subsection{Elastic properties}

The elastic properties of a solid are of importance. They are not only closely
related to various fundamental solid-state phenomena, such as atomic
bonding, equations of state and phonon spectra but are also
thermodynamically related to specific heat, thermal expansion, Debye temperature,
Gr\"{u}neisen parameter, and so on \cite{26}. Moreover, knowledge of the elastic
constants is very essential for many practical applications related to the
mechanical properties of a solid: such as load deflection, thermo-elastic
stress, internal strain, sound velocities, and fracture toughness.

In order to study the mechanical stability of TiN, we calculated the
second-order elastic constants $ C_{ij} $ from 0 to 20~GPa by using the
``stress-strain method'', which are listed in table~\ref{tab1}. The results of c-BN
and TiN at 0~GPa are close to the available data \cite{19, 28}. There are three
independent elastic constants $ C_{11} $, $ C_{12} $ and $ C_{44} $. The elastic
constant $ C_{11} $ represents the resistances to linear compression
at $ x $, $ y $ and $ z $ directions, the elastic constants $ C_{12} $ and $ C_{44} $ are
related to the elasticity in shape. For TiN, $ C_{11 } $ is the highest in all
the elastic constants, which implies that TiN is incompressible along the
coordinate axis.
\begin{table}[!h]
\caption{\label{tab1} The elastic constants $ C_{ij} $ of c-BN and TiN.}
\begin{center}
\renewcommand{\arraystretch}{0}
\tabcolsep=15pt
\begin{tabular}{|c||l|c|c|c|}
\hline\hline
& P (GPa) & $ C_{11} $ & $ C_{12} $ & $ C_{44} $ \strut\\
\hline
\rule{0pt}{2pt}&&&&\\
\hline
\raisebox{-2.7ex}[0pt][0pt]{c-BN}& 0& 779& 165& 446 \strut\\
\cline{2-5}
& 0 \cite{27} & 779& 165& 446 \strut\\
\cline{2-5}
& 0 \cite{28} & 798& 172& 469 \strut\\
\hline
\raisebox{-3.7ex}[0pt][0pt]{TiN}& 0& 581& 126& 166 \strut\\
\cline{2-5}
& 0 \cite{19} & 590& 145& 169 \strut\\
\cline{2-5}
& 10& 676& 141& 172 \strut\\
\cline{2-5}
& 20& 767& 155& 176 \strut\\
\hline\hline
\end{tabular}
\end{center}
\end{table}

\begin{figure}[!t]
\centerline{\includegraphics[width=\textwidth]{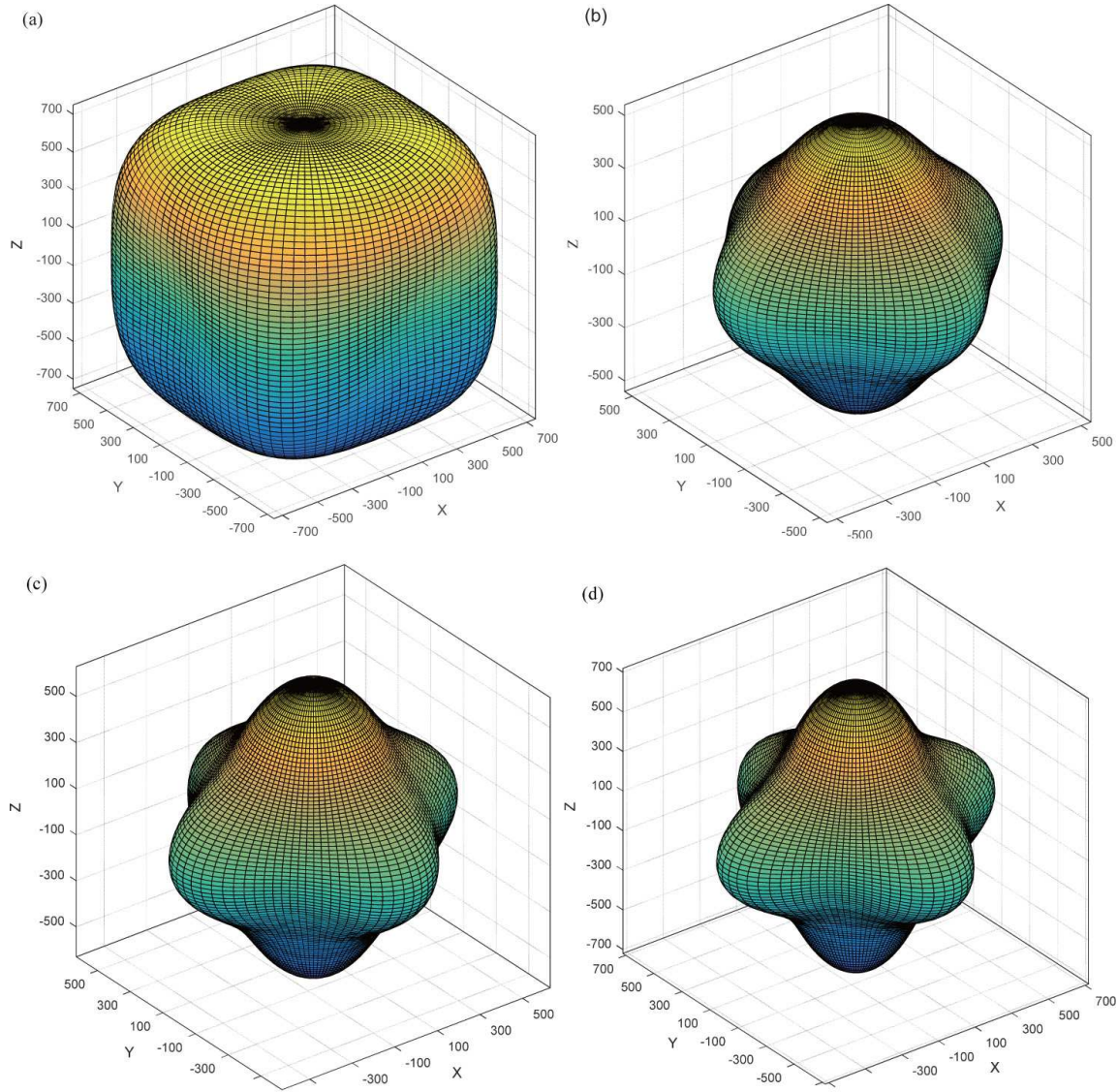}}
\caption{(Color online) The anisotropy of the Young's moduli for (a) c-BN at 0~GPa, (b) TiN
at 0~GPa, (c) TiN at 10~GPa and (d) TiN at 20~GPa.}
\label{fig3}
\end{figure}

The directional dependence of the Young's modulus of c-BN and TiN is
illustrated in figure~\ref{fig3}~(a)--(d). The 3D figures of the Young's moduli show the
deviation in shape from the sphere and the results indicate that the Young's
moduli are anisotropic. In addition, the Young's moduli maximal values of
c-BN and TiN at 0~GPa are 955 and 536~GPa, while the minimal values are 721
and 415~GPa. Thus, the ratios of the maximal and minimal values are 1.32 and
1.29. So, the anisotropy value of Young's modulus of TiN is smaller than
that of c-BN. At 10 and 20~GPa pressure, the Young's moduli maximal values
of TiN are 629 and 715, respectively, while the minimal values are 437 and
454. The anisotropy values are 1.44 and 1.58. Thus, the Young's moduli
anisotropy of TiN increases clearly with an increase of pressure.

\subsection{Thermodynamic properties}

Thermodynamic properties of TiN are determined in the temperature range
from 0 to 2000~K, meanwhile, the pressure effects are also analyzed in the
range from 0 to 20~GPa. By using the quasi-harmonic Debye model, thermal
expansion coefficients and volumes of TiN at a given pressure and
temperature conditions are calculated and the results are consistent with
 \cite{10}.

\begin{figure}[!t]
\centerline{\includegraphics[width=\textwidth]{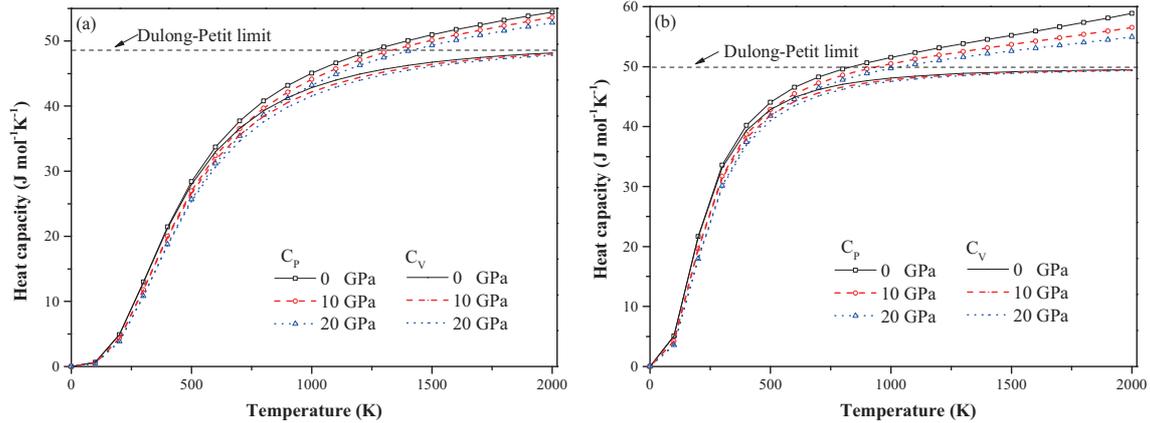}}
\caption{(Color online) Temperature dependence of the heat capacity at different pressures
for (a) c-BN and (b) TiN.}
\label{fig4}
\end{figure}

Heat capacity is an important parameter in condensed matter physics.
It provides an essential information on the heat transition process and
vibration properties. The relationships between isobaric heat capacity
$ C_{P} $ and constant volume heat capacity $ C_{V} $ versus temperature are
plotted in figure~\ref{fig4}. Our calculated $ C_{V} $ [in figure~\ref{fig4}~(b)] of TiN at 0~GPa is
consistent with the results in  \cite{10}. The difference between $ C_{P} $ and
$ C_{V} $ is very small, nearly below 500~K for c-BN and 400~K for TiN. Both
$ C_{P} $ and $ C_{V} $ increase rapidly with temperature due to the anharmonic
effect. At high temperature ranges, $ C_{V} $ approaches a constant value,
which is called a Dulong-Petit limit. The $ C_{P} $ still increases
monotonously with increments of temperature. Both $ C_{P} $ and $ C_{V} $ increase
with the temperature rise at a given pressure and slightly decrease with the
pressure rise at a given temperature. The effects of temperature on
the heat capacity are very significant. In addition, at lower temperatures,
the effects of temperature on TiN's heat capacity are stronger than those
of the temperature on c-BN's. At temperature below 1000~K, TiN's heat
capacity is larger than that of c-BN. Also, the Dulong-Petit limit of TiN is
a little larger than that of c-BN.

\begin{figure}[!t]
\centerline{\includegraphics[width=\textwidth]{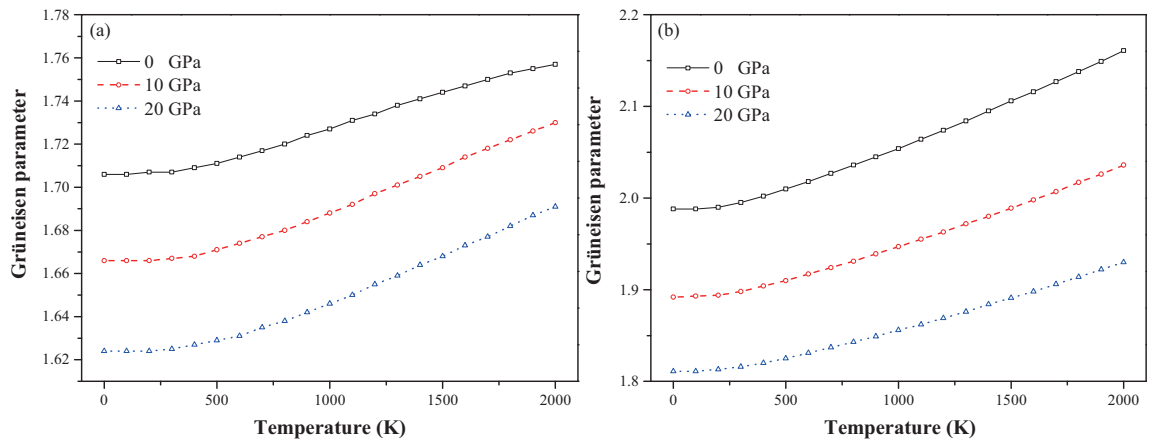}}
\caption{(Color online) The Gr\"{u}neisen parameter versus temperature of (a) c-BN and
(b)TiN.}
\label{fig5}
\end{figure}

The Gr\"{u}neisen parameter $ \gamma $ describes the variation in vibration
of a crystal lattice based on the changes of volume or temperature.
Recently, it has been widely used to characterize and extrapolate the
thermodynamic properties of materials at high pressures and high
temperatures, such as the temperature dependence of phonon frequencies and
line-widths. It is dominated by lower-frequency transverse modes at low
temperatures. The variation of Gr\"{u}neisen parameter with pressure and
temperature is displayed in figure~\ref{fig5}. The $ \gamma $ values rapidly decrease
with an increase of pressure and increase as temperature increment. The effect
of pressure and temperature on TiN's Gr\"{u}neisen parameter is more
significant than that of c-BN. The Gr\"{u}neisen parameter of TiN is larger
than that of c-BN.

\begin{figure}[!t]
\centering
\begin{minipage}{0.49\textwidth}
\begin{center}
\includegraphics{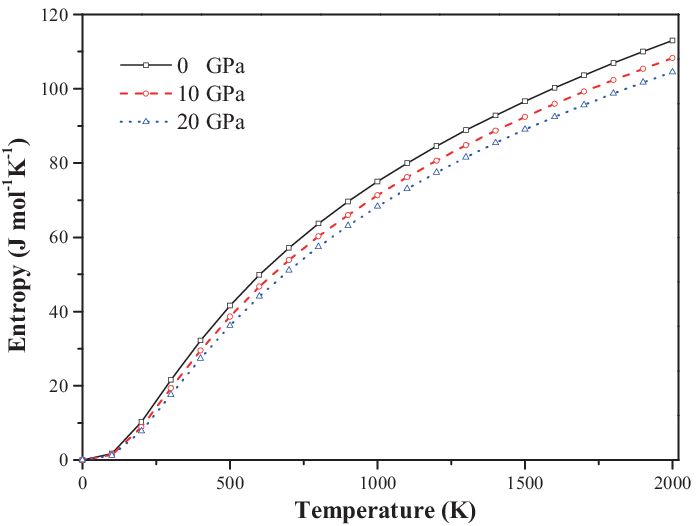}
\caption{(Color online) Variation of entropy with temperature at a given pressure for TiN.}
\label{fig6}
\end{center}
\end{minipage}
\begin{minipage}{0.49\textwidth}
\begin{center}
\includegraphics{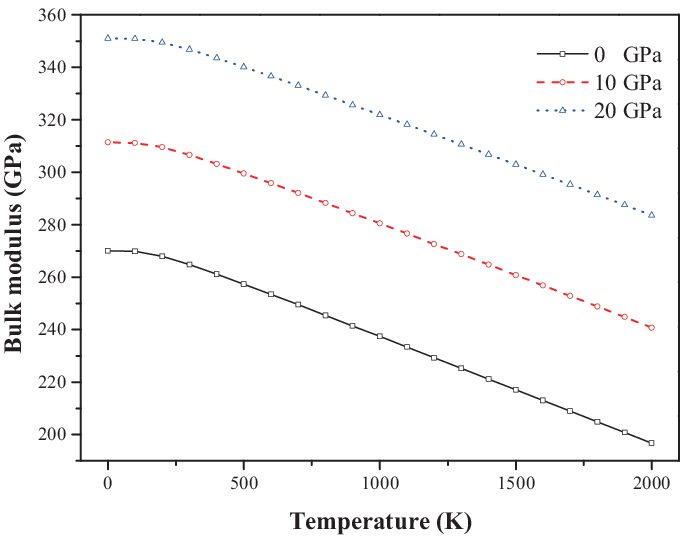}
\caption{(Color online) The bulk modulus versus temperature for TiN.}
\label{fig7}
\end{center}
\end{minipage}
\end{figure}

The entropy of a system is a measure of the amount of molecular disorder
within the system. A system can only generate, rather than destroy, entropy. The
entropy of a system can be increased or decreased by energy transports
across the system boundary \cite{29}. For TiN, the results of entropy as a
function of temperature are shown in figure~\ref{fig6}. It can be unambiguously seen
that entropy increases monotonously with an increasing temperature at a
given pressure and decreases with pressure rise at a given temperature. The
effect of temperature on entropy is obvious.

The bulk modulus, $ B $, can describe the elasticity of homogeneous isotropic
solids, which can be expressed as the force per unit area, and it indicates
the compressibility. The relationship between the bulk modulus and
temperature at different pressures is shown in figure~\ref{fig7}. These results indicate
that $ B $ decreases with temperature at a given pressure and rapidly increases
with pressure at a given temperature. The effects of pressure and
temperature on the bulk modulus are obvious.

\begin{figure}[!t]
\centering
\begin{minipage}{0.49\textwidth}
\begin{center}
\includegraphics{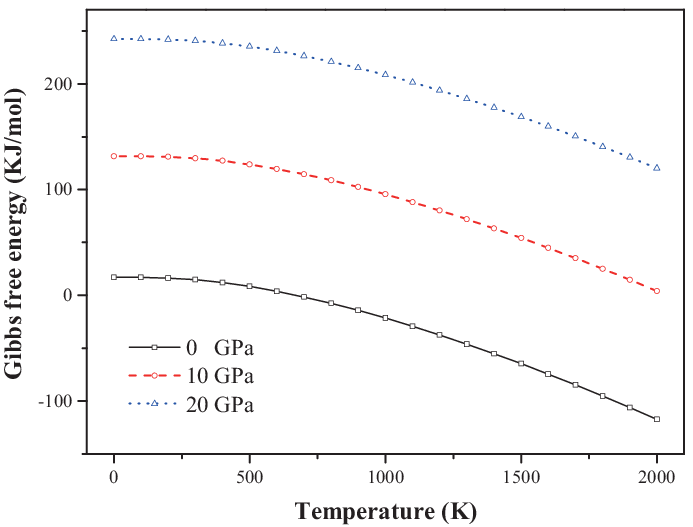}
\caption{(Color online) The Gibbs energy versus temperature for TiN at a given pressure.}
\label{fig8}
\end{center}
\end{minipage}
\begin{minipage}{0.49\textwidth}
\begin{center}
\includegraphics{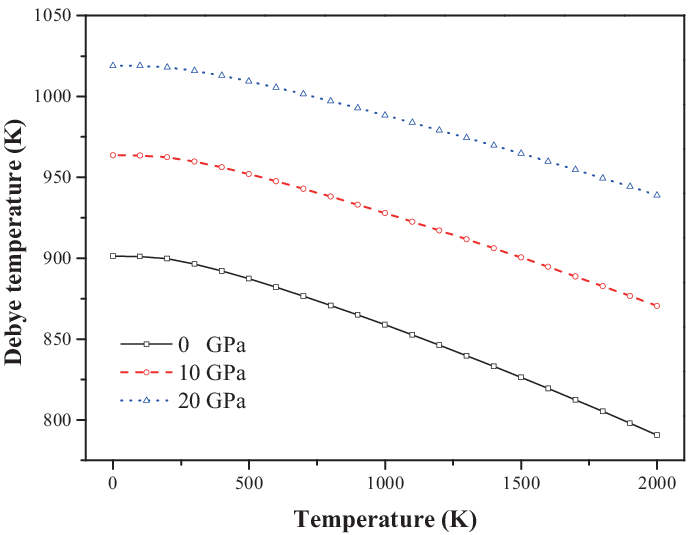}
\caption{(Color online) The Debye temperature versus temperature at various pressures for TiN.}
\label{fig9}
\end{center}
\end{minipage}
\end{figure}

Free energy is the inner energy in a thermodynamic system which can be
transformed into external work at a thermodynamic process. Gibbs free energy
is a kind of free energy. Figure~\ref{fig8} presents the relationships between the Gibbs
free energy and temperature at pressure range of 0${-}$20~GPa. The Gibbs free
energy is determined by enthalpy, entropy and temperature. The results show
that Gibbs free energy decreases with temperature rise. The effect of pressure
on the Gibbs free energy is also significant.

Debye temperature, $ \Theta_{\text D} $, is a cut-off on the~vibrational modes
introduced in the original Debye model. It can facilitate the calculation of the
integrals over the phonon density of states within a simple model where only
acoustic modes are kept but cut-off at some frequency so that the right
number of modes is obtained. It is one of fundamental parameters for solid
materials, which is correlated with many physical properties, such as
thermal expansion, melting point, and Gr\"{u}neisen parameter. It is used to
distinguish between high and low temperature regions for a solid. At low
temperatures, it is proportional to the sound velocity and directly related
to the elastic constants through bulk and shear modulus. As the temperature
above~$ \Theta_{\text D} $, we expect all modes to have the energy of $ \kappa T$ ($\kappa $ is Boltzmann's constant). At temperatures below it, one
expects the high-frequency modes to be frozen, i.e., vibrational
excitations arise solely from acoustic vibrations. Figure~\ref{fig9} shows the
dependence of the Debye temperature on temperature and pressure. It can be
seen that $ \Theta_{\text D} $ decreases with increasing temperature for TiN, at a
given pressure. At a given temperature, it rapidly increases with the applied
pressure rise. At zero pressure and zero temperature, the Debye temperature
value is equal to 901~K.

\section{Conclusion}

The static dielectric constants of TiN decrease with an increase of
pressure from 0 to 29~GPa, but it shows an anomalous behavior between 29 and
30~GPa, which is interpreted as a first order isostructural phase
transition. The absorption spectra exhibit high values ranging from the far
infrared region to the ultra-violet one. For TiN, $ C_{11} $ is the highest
in all the elastic constants, which implies that TiN is incompressible along
the coordinate axis. The Young's modulus anisotropy ratios are 1.29, 1.44
and 1.58 when the pressures are 0, 10 and 20~GPa, respectively. Thus, the
Young's modulus anisotropy of TiN clearly increases with the pressure rise. Thermodynamic properties of TiN are analyzed in detail by using the
quasi-harmonic Debye model under high temperature and pressure. The results
show that the effects of pressure and temperature on the bulk modulus,
Gr\"{u}neisen parameter, Gibbs free energy, and Debye temperature are
significant. The effect of temperature on the heat capacities and entropy
is obvious. At temperature below 1000~K, TiN's heat capacity is larger than
that of c-BN. The Gr\"{u}neisen parameter of TiN is larger than that of
c-BN.

\section*{Acknowledgements}

This work was supported by Natural Science Basic Research plan in Shaanxi
Province of China (Grant No.~2016JM1026) and the 111 Project (B17035).

\ukrainianpart

\title{Дослідження з перших принципів оптичних,  пружних анізотропних і термодинамічних властивостей   TiN при високих температурах і тисках%
}
\author[R. Yang, C. Zhu, Q. Wei, K. Xiao, Z. Du]{Р. Янг\refaddr{label1}, Ц. Жу\refaddr{label1}, К. Вей\refaddr{label1}, K. Шіао\refaddr{label1}, Ж. Ду\refaddr{label2}}
\addresses{
\addr{label1} Школа фізики та оптоелектронної інженерії, університет округу Шідіан, Сіань, Шенсі 710071, Китай
\addr{label2}  Національний центр високопродуктивних обчислень, м. Шеньчжень  518055, Китай
}

\makeukrtitle

\begin{abstract}
Детально проаналізовано оптичні, пружні анізотропні та термодинамічні властивості TiN в структурі
NaCl (B1) в температурній області від 0 до  2000~K і в діапазоні тисків від 0 до 20~GPa. З обчислених діелектричних постійних знайдено ізоструктурний фазовий перехід першого роду між
29 і 30~GPa. Спектр поглинання демонструє високі значення від інфрачервоної області до ультрафіолетової. Значення анізотропії модуля Юнга для  TiN є менше ніж для  c-BN при 0~GPa, і анізотропія  TiN чітко зростає з ростом тиску. Впливи тиску і температури на об'ємний модуль, параметр Грюнайзена, вільну енергію Гіббса і температуру Дебая є значними. Параметр Грюнайзена для TiN є набагато більшим ніж для  c-BN. При температурах нижчих  1000~K, питома теплоємність TiN є набагато більша ніж c-BN.
\keywords TiN, оптичні властивості, пружна анізотропія, термодинамічні властивості, перші принципи
\end{abstract}

\end{document}